\documentclass[manuscript]{aastex}
\usepackage{soul,color}

\slugcomment{}

\shorttitle{HFQPO Studies with X-ray Observations}
\shortauthors{Beheshtipour, Hoormann, \& Krawczynski}

\begin{document}

\title{Studies of the Origin of High-Frequency Quasi-Periodic Oscillations of Mass Accreting Black Holes in X-ray Binaries 
with Next-Generation X-ray Telescopes}

\author{Banafsheh Beheshtipour\altaffilmark{1}, Janie K. Hoormann, and Henric Krawczynski}

\affil{Physics Department and McDonnell Center for the Space Sciences, Washington University in St. Louis, One Brookings Drive, CB 1105, St. Louis, MO 63130, USA}

\altaffiltext{1}{Email: b.beheshtipour@wustl.edu}

\begin{abstract}
Observations with {\it RXTE} (Rossi X-ray Timing Explorer) revealed the presence of High Frequency
Quasi-Periodic Oscillations (HFQPOs) of the X-ray flux from several accreting stellar mass Black Holes.
HFQPOs (and their counterparts at lower frequencies) may allow us to study general relativity in the strong
gravity regime. However, the observational evidence today does not yet allow us to distinguish between 
different HFQPO models. In this paper we use a general relativistic ray-tracing code to investigate 
 X-ray timing-spectroscopy and polarization properties of HFQPOs in the orbiting Hotspot model. 
We study observational signatures for the particular case of the 166 Hz quasi-periodic oscillation (QPO) 
in the galactic binary GRS~1915+105.  We conclude with a discussion of the observability of spectral signatures 
with a timing-spectroscopy experiment like the {\it LOFT} (Large Observatory for X-ray Timing) 
and polarization signatures with space-borne X-ray polarimeters such as {\it IXPE} (Imaging X-ray Polarimetry Explorer), {\it PolSTAR} (Polarization Spectroscopic Telescope Array), {\it PRAXyS} (Polarimetry of Relativistic X-ray Sources), or {\it XIPE} (X-ray Imaging Polarimetry Explorer). 
A high count-rate mission like {\it LOFT} would make it possible to get a QPO phase for each photon, enabling
the study of the QPO-phase-resolved spectral shape and the correlation between this and the flux level.
Owing to the short periods of the HFQPOs, first-generation X-ray polarimeters would not be able to assign a 
QPO phase to each photon. The study of QPO-phase-resolved polarization energy spectra would thus
require simultaneous observations with a first-generation X-ray polarimeter and a {\it LOFT}-type mission.
\end{abstract}


\section{Introduction}
The X-ray observations of accreting neutron stars and black holes (BHs) of the last one and a half decade have 
revealed new avenues for testing General Relativity (GR) in the strong gravity regime \citep{sch04,psa08}. 
The {\it RXTE} revealed High Frequency ($>$40~Hz) QPOs 
in a number of accreting BHs in X-ray binaries \citep{rem06}. Altogether, HFQPOs have been found in seven
systems, three with a detection at a single frequency, and four with a detection at multiple frequencies.
The binaries GRO~J1655-40 and possibly also GRS1915+105 exhibit pairs of HFQPOs with frequencies 
at a 3:2 frequency ratio \citep{rem06}. 
QPOs may become a powerful tool for the study of BHs, i.e. to inform us about the emission state,
the BH spin, and/or to test GR in the strong gravity regime. 

A number of models have been developed 
to explain the observed QPOs in different frequency ranges. \citet{ste98} explain the HFQPOs in
Low Mass X-ray Binaries (LMXBs) as the general relativistic Lense-Thirring precession of the innermost disk region.
\citet{abr01} explain HFQPOs as a resonance of the orbital and epicyclic motion of the accreting matter.
\citet{bur05} shows that the resonance between the vertical epicyclic frequency and the periastron precession frequency gives for the source GRO J1655-40 a spin estimate which is consistent with that from the X-ray continuum method. The torus model, first presented by  \citet{rez03}, posits that p-mode oscillations of an accretion torus cause the HFQPOs. \citet{bur04} study the flux variability induced by radial torus oscillations for a Schwarzschild BH. Their calculations indicate that the high-frequency modulation of the X-ray flux could result from light bending in the strongly curved BH spacetime (causing high-frequency flux variations). The resonance model of \citet{pet08} explains HFQPOs as resulting from the resonance of a spiral wave in the inner part of the accretion disk with vertical epicyclic oscillations. Recently, \citet{dex14} proposed that the local and vertical epicyclic and acoustic breathing modes could lead to observing HFQPOs in the steep power law spectral state. \citet{wag01} and \citet{kat03} also explain HFQPO with the adiabatic perturbations of the relativistic accretion disk.
Last but not least, \citet{tag06} and \citet{kei08} explain the HFQPOs as the observational signature of magnetohydrodynamic Rossby wave instabilities (RWI) and as the light echo, respectively. 

In this paper we study observational signatures predicted by the hotspot (HS) model \citep{sch04}. 
This model assumes an accretion disk with a bright HS orbiting the black hole. 
The model is motivated by the similarity between the HFQPO and the coordinate frequency near the ISCO. 
Furthermore, following the work of \citet{mer99}, the resonance between azimuthal and radial oscillations may explain the observed integer commensurabilities between different HFQPO frequencies \citep{sch04}. Also, recently \citet{li14} argued that the HS model can distinguish BHs and wormholes based on infrared observations.

In this paper we study the spectral and spectropolarimetric observational signatures of the HS model.
A timing and spectroscopy mission like {\it LOFT} is ideally suited to detect HFQPOs \citep{boz13,fer12,vin13} and
to measure QPO-phase-resolved energy spectra. Spectroscopic X-ray polarimetry observations  
\citep[see e.g.][]{mes88, lei97, bel10, kra12,li09, sch09}, offering three times as much information as purely 
spectroscopic observations (i.e. the Stokes parameters $I$, $Q$ and $U$ rather than $I$ alone
as function of energy) would offer additional handles to distinguish between HFQPO models. 
As some X-rays scatter before leaving the accretion disk, even the thermal emission is polarized 
(e.g. \citet{li09} and references therein). The polarization angle changes as the X-rays propagate 
through the strongly curved spacetime of the BH. Additional photon scattering off the accretion disk 
or in the accretion disk corona modifies the polarization fraction and angle. 
\citet{zam09,zam11} studied the polarization of HSs orbiting supermassive BHs at infrared wavelengths
and used infrared observations of the supermassive BH Sgr $A^*$ to constrain its mass and spin.

We use the ray-tracing code developed by \citet{kra12} to model the X-ray emission from HSs orbiting 
Schwarzschild and Kerr stellar mass BHs in X-ray binaries. Although our studies are generic in nature, 
our Kerr black hole calculations adopt parameters chosen to describe the 166\,Hz QPO of the galactic BH GRS 1915+105. 
HFQPOs has been observed in steep power law state (SPL) of BHs.
The SPL state commonly attributed to a corona of hotter gas which reprocesses the accretion disk photons and gives rise to a power law emission spectrum \citep[e.g.][]{rem06}. Since the geometry and physical properties of a corona are not fully understood, a wide range of coronal models have been proposed \citep[e.g.][]{haa91,dov97,now02,mcc03,sch10}.
In this paper we model a geometrically thin, optically thick accretion disk with an orbiting hotspot with and without a sandwich corona. The corona properties have been chosen to give the power law energy spectrum of GRS 1915+105 in the SPL state.

In a somewhat related study \citet{ing15} have studied the polarization properties of {\it low frequency QPOs} 
assuming that they originate from the Lense-Thirring precession of the inner accreting flow. 
They find polarization fraction variations on the order of 1\% which could be detected and 
studied by an X-ray polarimeter with hard X-ray sensitivity such as the proposed {\it PolSTAR} mission.

The rest of the paper is structured as follows. We summarize the  HS model and describe our simulations in Sect.~2. Section~3 presents the results for  Schwarzschild and Kerr BHs, including a discussion of the observational signatures as function of the HS parameters. In Sect.~4 we summarize the results and discuss the expected energy spectra and polarization signatures of competing HS models. Throughout this paper, all distances are in units of gravitational radius $r_{\rm g} =GM/c^2$, 
and we set $G=c=\hbar=1$. 

\section {Methodology}
\subsection{The hotspot model}
\citet{ste98,ste99}  introduced the HS model to explain the observations of  QPOs with frequencies comparable to the orbital frequencies of matter orbiting BHs and neutron stars close to the ISCO. As mentioned above, the HS can explain not only the detection of HFQPO at one frequency but also twin HFQPOs with integer frequency ratios as the result of non-linear resonances occurring near geodesic orbits  \citep{abr01,abr03a}.

The HS model posits that a region with a temperature exceeding that of the ambient material orbits the BH. We assume that the all the matter orbits the BH on a nearly circular orbit around the BH with the angular frequency $\nu_\phi$ given by \citet{bar72}:
\begin{equation}
\Omega_\phi=2 \pi \nu_\phi=\frac{\pm \sqrt{M}}{r^{3/2} \pm a \sqrt{M}}.
\end{equation}
 For a prograde (retrograde) orbit, the upper (lower) sign applies. Typically, we consider HSs with a radius of around $0.25-0.5$ $r_{\rm g}$. It has been argued that a larger HS will not survive a long time because of the viscous shearing of the disk \citep{mar00}. \citet{sch04} have shown that the light curve and the HFQPO power spectrum are independent of the HS's size and shape. They also tried to explain the 3:2 commensurability for twin peaks in some X-ray binary systems with the idea of a noncircular orbit of the HS and its different coordinate frequencies. These properties lead to some beat frequency in the light curve and they believe that one of the peaks is at the azimuthal frequency and the other is at beat modes $\nu_\phi \pm \nu_r$.

\citet{sch05} modeled HFQPOs with a couple of orbiting HSs assuming a random phase, different life times, and a finite width for  HSs. The model produced broad HFQPOs with {\it Q}-factors, defined as the ratios of the HFQPO line centroids to the line widths (FWHM), matching the observed ones.  

We neglect the Faraday rotation that polarized photons would experience in a magnetized plasma, an assumption which
seems to hold at photon energies exceeding a few keV \citep{dav09}.
\subsection{Thermal disk simulation}
We assume that the HS is a disk segment emitting with a temperature five times higher than the surrounding material. This temperature gives (for the adopted hotspot size) HFQPO rms amplitudes comparable to the observed ones. The effects of the hotspot size on the observable signatures are discussed in the result section.
The HS of the Schwarzschild BH extends from the innermost stable circular orbit ($r_{\rm ISCO}=6$) 
to $r=r_{\rm ISCO}+2 \Delta r$, where $\Delta r=0.5$ and from $\phi$ to $\phi+  \Delta \phi$, where $\Delta \phi=0.08\pi$. 
The HS of the Kerr BH is centered at the radial coordinate $r=5+\Delta r$ to model the 166\,Hz QPO of GRS 1915+105.

We use the  general relativistic ray-tracing code of \citet{kra12}. Photons are tracked forward in time from their emission site to the observer, including, if applicable, one or several scatterings off the accretion disk. The standard Novikov-Thorne radial brightness profile of a geometrically thin, optically thick accretion disk is used to weigh the simulated rays \citep{nov73, pag74}. Recent General Relativistic Magnetohydrodynamic (GRMHD) simulations show that the Novikov-Thorne results are a good approximation of the more detailed results \citep{nob09, pen10, pen12}.

The code simulate an accretion disk extending  from $r_{\rm ISCO}$ to $r_{\rm max}=100 r_{\rm g}$.
For each background metric, we divide the accretion disk in 10,000 radial bins spaced equally in the logarithm of the Boyer Lindquist coordinate r. For each radial bin, we  simulate 1000 photon packages. For the radially symmetric accretion disk emission, the code makes use of the azimuthal symmetry of the problem: all photons are launched at an azimuthal angle $\phi=0$. When they leave the simulation sphere, we infer that the probability to find them in the azimuth angle interval from $\phi$ to $\phi+\Delta \phi$ equals $\Delta \phi/2\pi$. Photons are created in the plasma frame  with a limb darkening function from \citet{cha60}. The code uses Table \textrm{XXIV} of \citet{cha60} to calculate the initial polarization of the photon and the statistical weight for its emission direction. Subsequently, the photon wave vector and polarization vector are transformed into the Boyer-Lindquist frame. The photons are then tracked by solving the geodesic equation:
\begin{equation}
\frac{d^2 x^\mu}{d \lambda ^{\prime 2}}= -\Gamma^\mu_{ \;\;\sigma \nu} \frac{d x^\sigma}{d \lambda ^\prime} \frac{d x^\nu}{d \lambda ^ \prime},
\end{equation}
with $\lambda^\prime$ being an affine parameter and $\Gamma^\mu_{\;\;\sigma \nu}$ the Christoffel symbols. The polarization vector is parallel transported with the equation
\begin{equation}
\frac{d f^\mu}{d \lambda ^{\prime}}= -\Gamma^\mu_{ \;\;\sigma \nu} f^\sigma \frac{d x^\nu}{d \lambda ^ \prime},
\end{equation}
 At the analysis stage, the photon packages are weighted to mimic the radial brightness distribution $F(r)$ of \citet{pag74}. The latter authors used the conservation of mass, angular momentum, and energy to derive
\begin{equation}
F(r)=\frac{-\dot{M}}{4 \pi}e^{-(\nu+\psi+\mu)} \frac{p^t_{,r}}{p_\phi}\int\limits_{r_{ISCO}}^{r}\frac{p_{\phi,r}}{p^t}dr
\end{equation}
where $\dot{M}$ is the accretion rate  for a stationary, axially symmetric metric given by the functions $\nu$, $\psi$, and $\mu$, and $p^{\mu}$ is the four-momentum of the disk material and ``,'' denotes the ordinary partial differentiation (see \citet{bar72} and \citet{pag74} for the nomenclature). Photon packages are assumed to be emitted with a Blackbody energy spectrum with the temperature
\begin{equation}
T_{eff}=(\frac{F(r)}{\sigma_{SB}})^{\frac{1}{4}}
\end{equation}
The code tracks the red-and blue-shifts of the photons between emission and absorption (including the gravitational redshifts and blueshifts incurred during propagation), and the thermal energy spectrum is then red- or blue-shifted when accounting for the detected photon packages. The HS is treated in the same way, except that the effective temperature and thus the brightness and the statistical weight is higher for this segment. For simplicity we do not reduce the temperature of the adjacent parts of the accretion disk which would be required in a self-consistent steady-state solution. The slight temperature reduction of the adjacent material would enlarge the contrast between the hotspot and the disk and would thus enlarge the observational signatures.

When a photon hits the accretion disk, it is scattered into a random direction with equal probability in solid angle and with a statistical weight determined from Table \textrm{XXIV} of \citet{cha60}. 
The photon is tracked until it comes too close to the event horizon ($r<r_H+0.02$), or until its radial Boyer Lindquist coordinate $r$ exceeds 10,000 $r_{\rm g}$. If the latter happens, the photon is back-tracked to $r=10,000$. Subsequently, its wave and polarization vectors are transformed into the reference frame of an observer at fixed coordinates. These results, together with the information about the polarization degree (a Lorentz invariant) 
are then used to determine the photon energy and the Stokes parameters  $I$, $Q$, and $U$ of all photons
detected in a certain polar angle range. In the final step, the stokes parameters are used to find the polarization fraction and angle of each photon.
The interested reader can find a more detailed description of the ray-tracing code in \citet{kra12}. 
\subsection{Corona simulation}
\begin{figure}
\epsscale{0.8}
\plotone{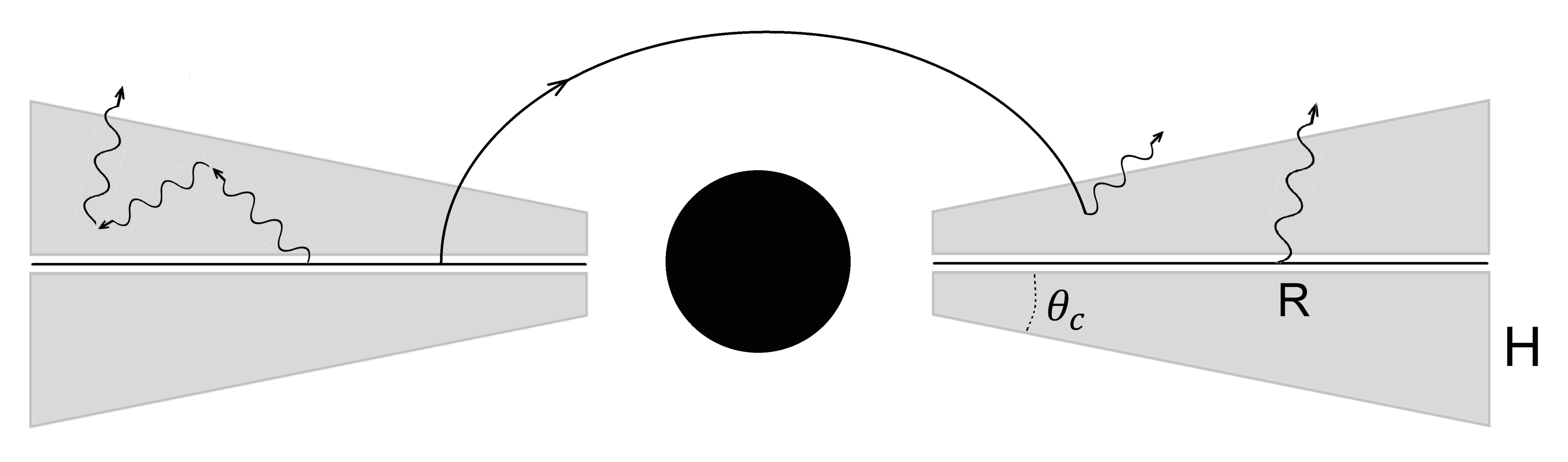}  
\caption{Sketch of a wedge corona geometry and some possible photon paths. The corona extends above and below the accretion disk with a constant opening angle, $tan(\theta_{c})=H/R$. Photons may reach the observer directly or scatter once or multiple times in the corona and/or off the disk. The strong gravitational field also deflects the photon paths.
}\label{wedgecorona}
\end{figure}

We simulate a simple Corona geometry, an isothermal layer with a constant opening angle forming a wedge above and below the accretion disk (see Figure~\ref{wedgecorona}). The vertical optical depth of this layer is set to a constant, $\tau_{0}=0.2$ and the temperature of the hot electrons in the corona is set to $T_{corona}=30$ keV. These parameters reproduce the observed photon index for GRS 1915+105 in the SPL state \citep{bel06}. The opening angle of the wedge is set to 2$^{\circ}$. A larger opening angle would result in longer light travel times inside the corona and would result in a wider X-Ray pulse from the hotspot. We assume that the corona gas orbits the black hole with the angular velocity of a zero angular momentum observer \citep[ZAMO,][]{bar72}. As we track individual photons originating from the accretion disk, we check for each integration step if the photon is inside the corona. If so, we transform the start and end point of the integration step into the rest frame of the corona plasma, and determine the optical depth between these two points. The optical depth is then used to determine the probability for Thomson scattering. We have implemented the scattering in both, in the Thomson and the Klein-Nishina regimes, but we use here the algorithm for Thomson scatterings (see the discussion of \citep{sch10} for a justification). We then draw a random direction of the scattering electron in the comoving plasma rest frame, transform the wave vector of the photon from the plasma rest frame into the rest frame of the electron, and determine the photon wave vector after scattering. Subsequently, the photon wave vector is first transformed back into the rest frame of the corona, and subsequently, into the global Boyer Lindquist coordinates. A more detailed description of the modeling of the Comptonization of the photons in the corona will be given in a companion paper (Beheshtipour et al., in preparation).
\section{Results}
In the following, all results will be given for the $2-30$ keV energy band unless otherwise specified. The inclination is $i=0^{\circ}$ 
for an observer viewing the disk face-on and $i=90^{\circ}$ for an observer viewing the disk edge-on. In all light curves (intensity plots) we give the average $2-30$ keV photon flux.
\subsection{Thermal emission}
\begin{figure}
\epsscale{0.8}
\plotone{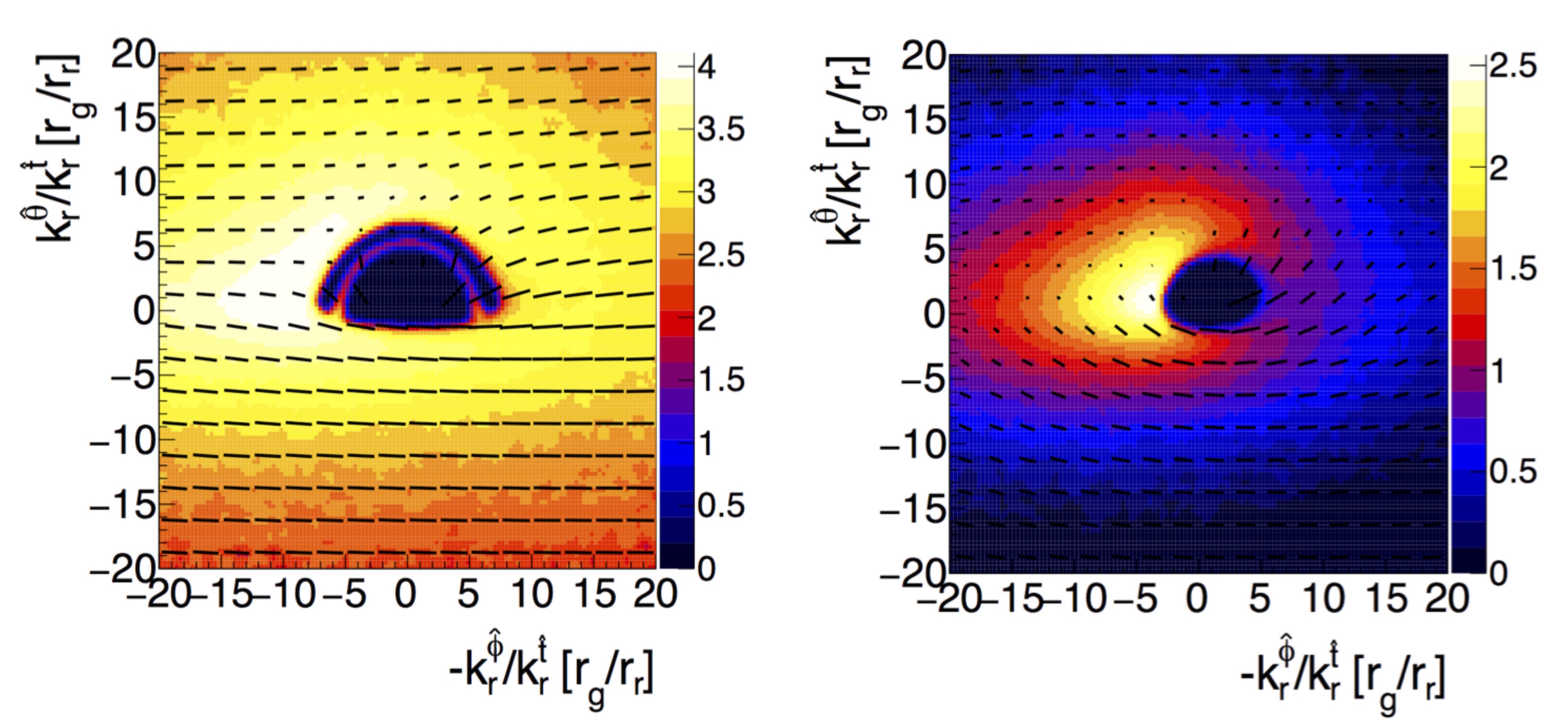}  
\caption{Image of the steady emission from the accretion disk of a Schwarzschild BH ($a=0$ and $M=10 M_{\sun}$), at an inclination of $75^{\circ}$ (left panel), and GRS~1915+105 ($a=0.95$ and $M=14 M_{\sun}$), at an inclination of $66^{\circ}$ (right panel). The observed intensity is color-coded on a logarithmic scale. The length and orientation show the polarization fraction and polarization angle, respectively. We measure the polarization angle from the projection of the spin axis of the BH in the plane of the sky and it increases for a clockwise rotation when looking toward the BH.
\newline(A color version of this figure is available in the online journal.)
}\label{map}
\end{figure}

Figure~\ref{map} shows an accreting Schwarzschild BH at an inclination of $75^{\circ}$, and the accreting Kerr black (GRS~1915+105) seen at an inclination of 66\degr\ \citep{fen99} for 2-15 keV energy band. The lengths and orientations of the bars in the image show the polarization fractions and angles, respectively.  The image clearly shows the relativistic beaming and de-beaming off the emission from the disk resulting in pronounced brightness variations across the disk \citep[see also][]{sch09}.

\begin{figure}
\epsscale{0.8}
\plotone{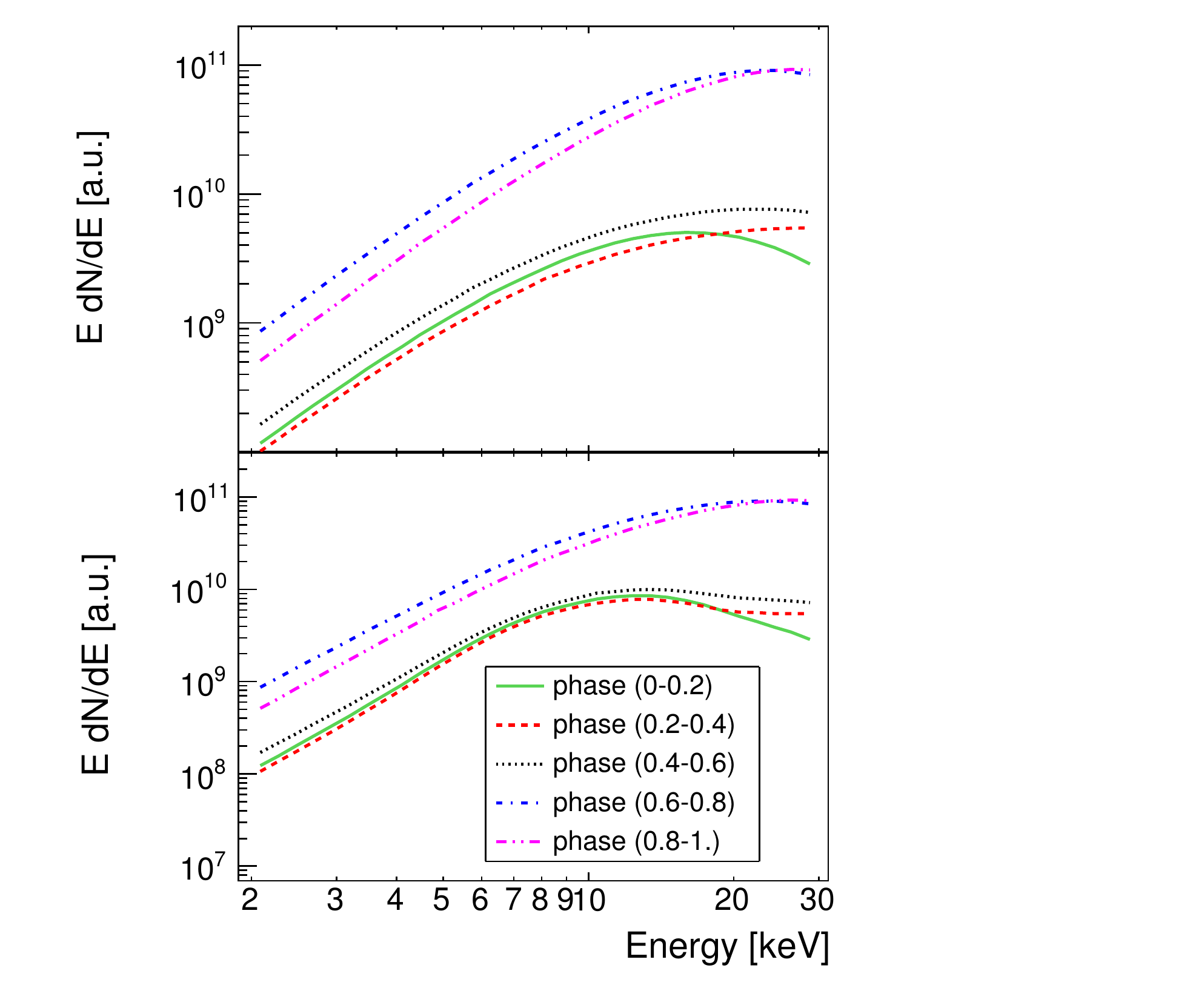} 
\caption{Phase resolved energy spectra of a HS orbiting GRS~1915+105 for HS emission (upper panel) and total emission (lower panel). 
\newline(A color version of this figure is available in the online journal.)
} \label{flux}
\end{figure}

\begin{figure}
\epsscale{0.8}
\plotone{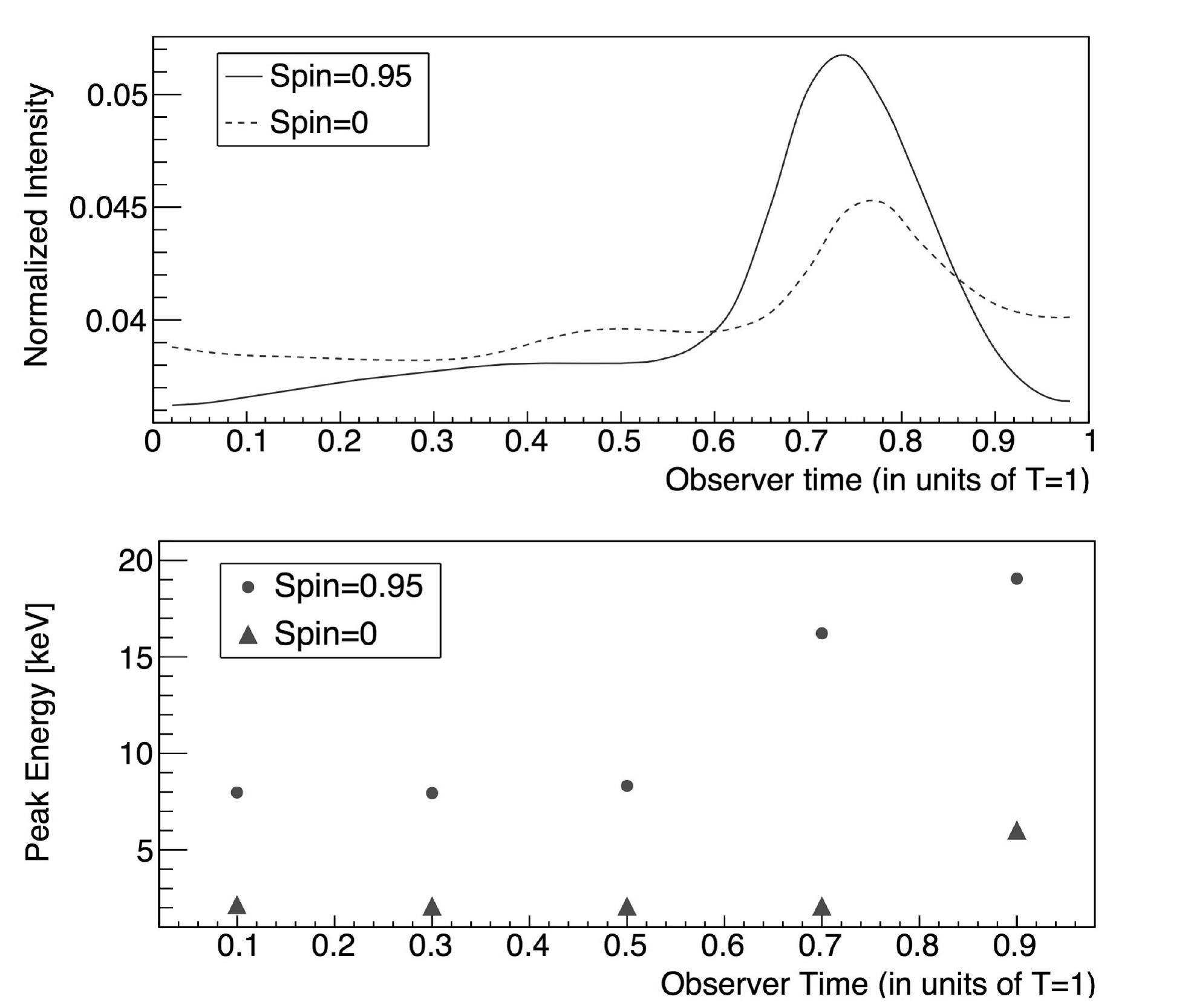}
\caption{Intensity (upper panel), average $2-30$ keV photon flux, and peak energy (lower panel) of the total X-ray emission from the Schwarzschild and Kerr  BHs. The intensity is normalized to 1 when integrated over all phases.} \label{fluxinten}
\end{figure}

We plot the phase resolved energy spectra of the accretion disk and HS emission of the GRS 1915+105 model in Fig.~\mbox{\ref{flux}}.
We divided the orbit into 5 phase bins and each line in the figure shows the energy spectra of the HS 
for the specific phase bins. 
Note that the phase also characterizes the azimuthal position of the HS. At phase~=~0 (0.5) the HS is closest to (furthest away from) the observer.  The energy spectra exhibit well defined flux peaks. 
The corresponding energy to these peaks, i.e. peak energies, for the total emission (HS plus accretion disk) are shown in Fig.~\ref{fluxinten} as function of phase for both simulated BHs. 
The peak energies are higher for the Kerr black hole as its HS is closer to the BH (the HS center is at 5.5~$r_g$ for the Kerr BH and 6.5 $r_g$ for Schwarzschild BH) allowing for bright emission from the inner regions of the accretion disk.
The integral flux, intensity, drops in the last phase bin even though the energy spectrum still hardens (Fig~\ref{flux}), owing to the Doppler shift from the relativistic motion of the accretion disk plasma. This can be understood as follows: Photons returning to the disk and scattering off the disk have a very broad energy spectrum owing to the energy gains/losses incurred during the scattering process. These scattered photons come a bit later than the unscattered photons, giving rise to the hard spectrum at the end of the peak. Interestingly, the flux peak leads the peak of the spectral hardness by $\sim$~0.2 in phase.

\begin{figure}
\epsscale{1}
\plotone{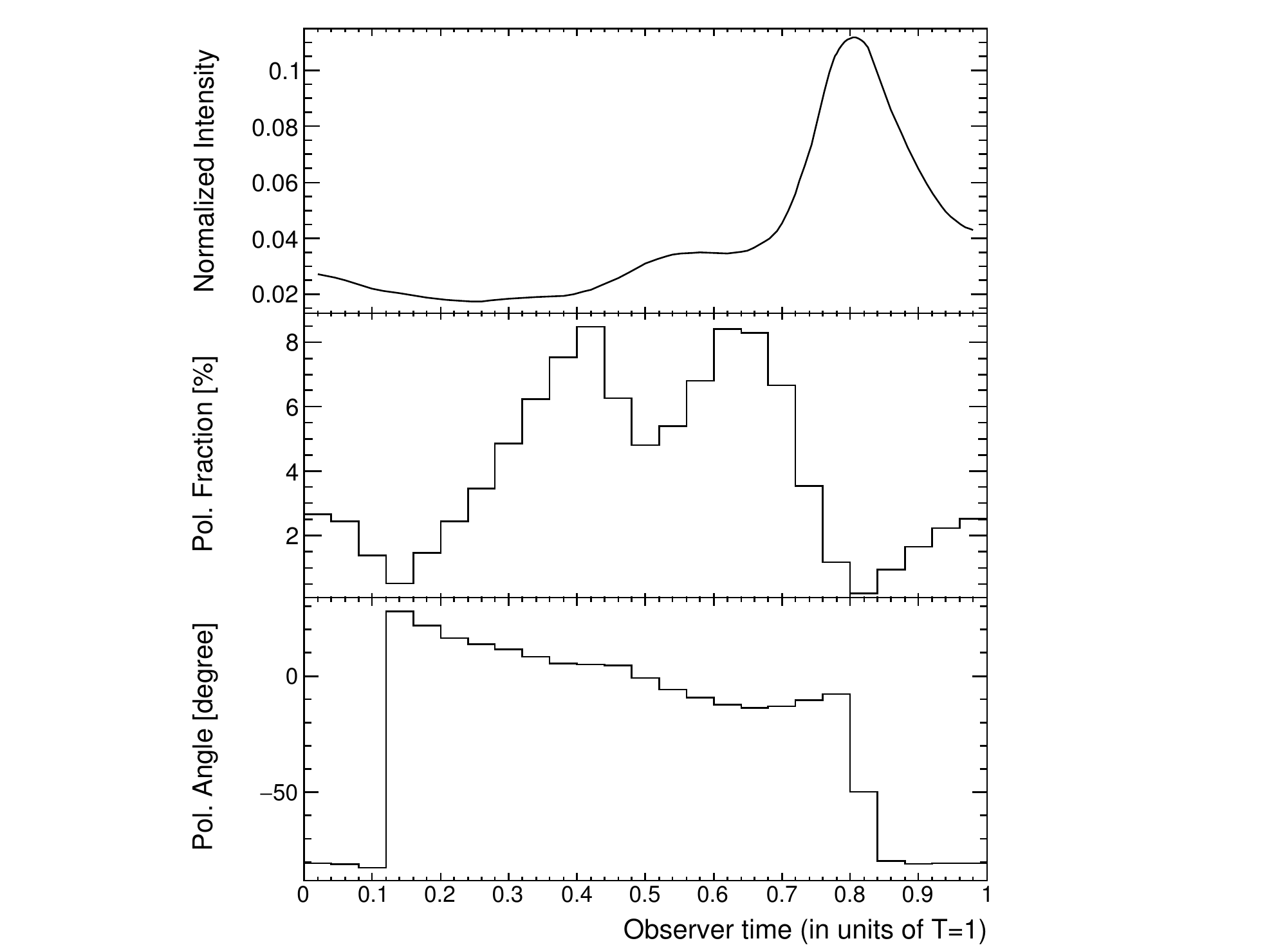}
\caption{Intensity, polarization fraction, and polarization angle of the HS emission for a Schwarzschild BH, viewed at an inclination of $75^{\circ}$. The emission is polarized with a max polarization fraction of $\approx$ 8.5\%. The polarization angle exhibits a full $180^{\circ}$ swing in one orbit. A polarization angle of 0$^{\circ}$ corresponds to emission with an electric field vector 
perpendicular to the accretion disk spin axis.}\label{Ipol0}
\end{figure}

\begin{figure}
\epsscale{1}
\plotone{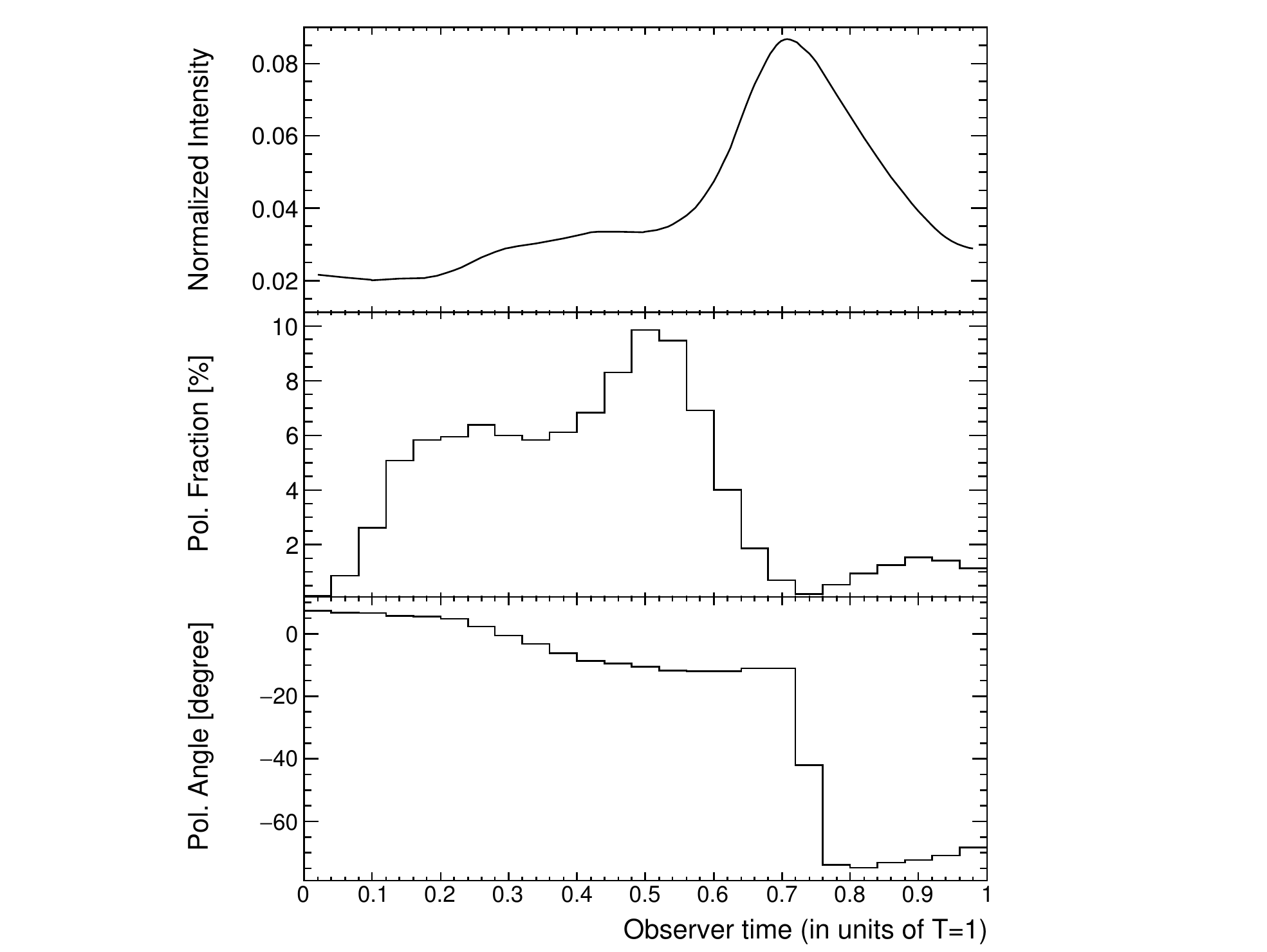}
\caption{The same as Fig.~\ref{Ipol0} for a HS orbiting the Kerr BH. The emission is highly polarized with a max polarization fraction of $\approx$ 10\%. The polarization angle swings by $90^{\circ}$ during one orbit.}\label{Ipol9}
\end{figure}

\begin{figure}
\epsscale{0.8}
\plotone{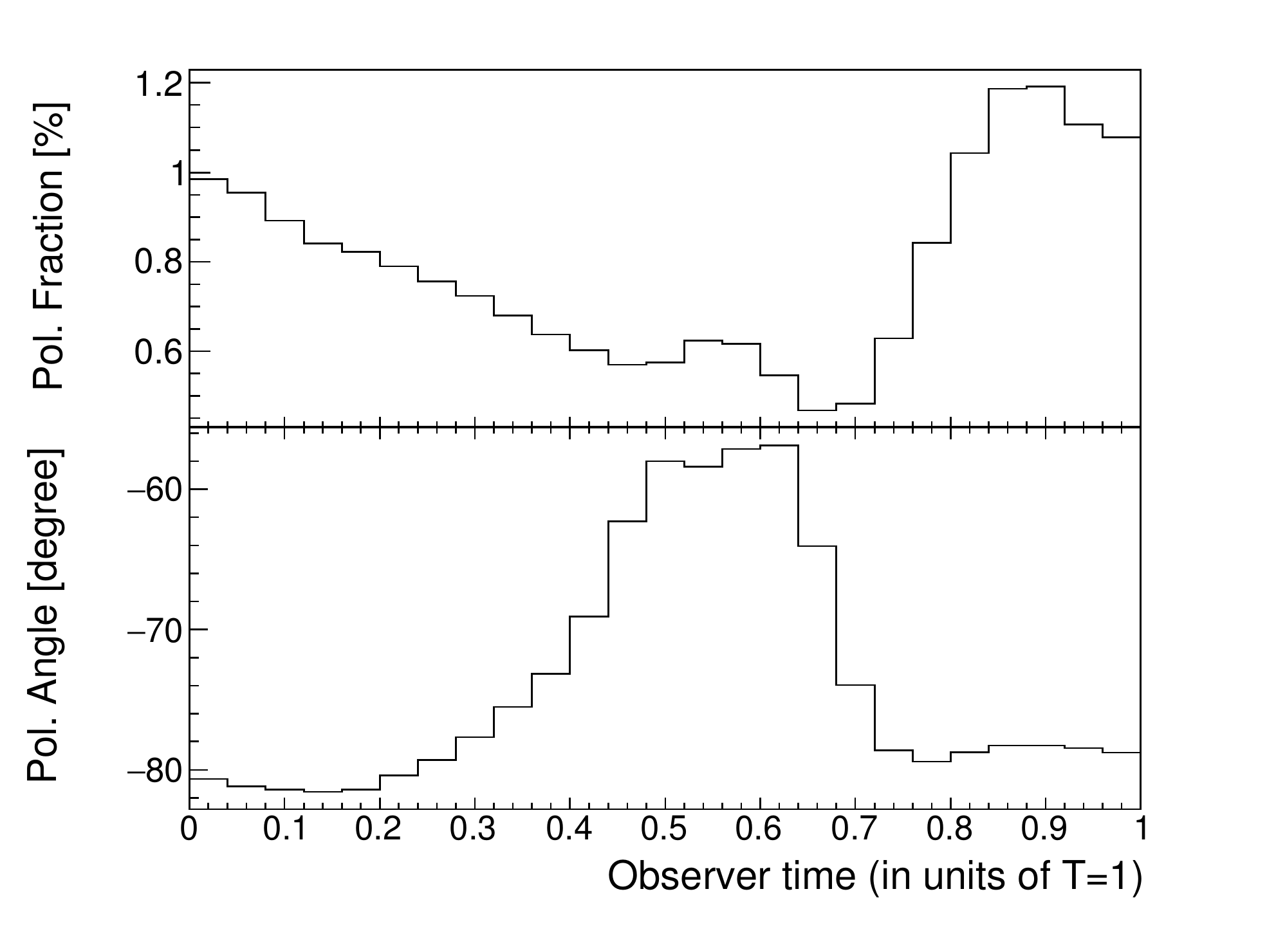}
\caption{Polarization fraction and angle of the HS plus disk emission for the Kerr BH. This is polarized with a max polarization fraction of $\approx$ 1.2\%.}\label{totalpol}
\end{figure}

Figure~\ref{Ipol0} and~\ref{Ipol9} show the normalized intensity, polarization fraction and angle for the Schwarzschild and Kerr BHs. The intensity is normalized to 1 when integrated over all phases. Comparing these figures one can see the effect of the BH spin on the polarization of the observed emission. 
Interestingly, the HS model predicts that the peak of the emission (dominated by direct HS emission relativistically 
beamed towards the observer) is accompanied by a drop in polarization fraction  and a large swing of the polarization direction. 
As shown below, the polarization properties result from the competition of the direct HS emission and the HS emission 
reflecting off the accretion disk. Also, the effect of including the emission of both the disk and HS on the polarization fraction and angle is shown in figure~\ref{totalpol}. The total polarization is lower due to the disk emission being less polarized.

\begin{figure}
\epsscale{1.1}
\plotone{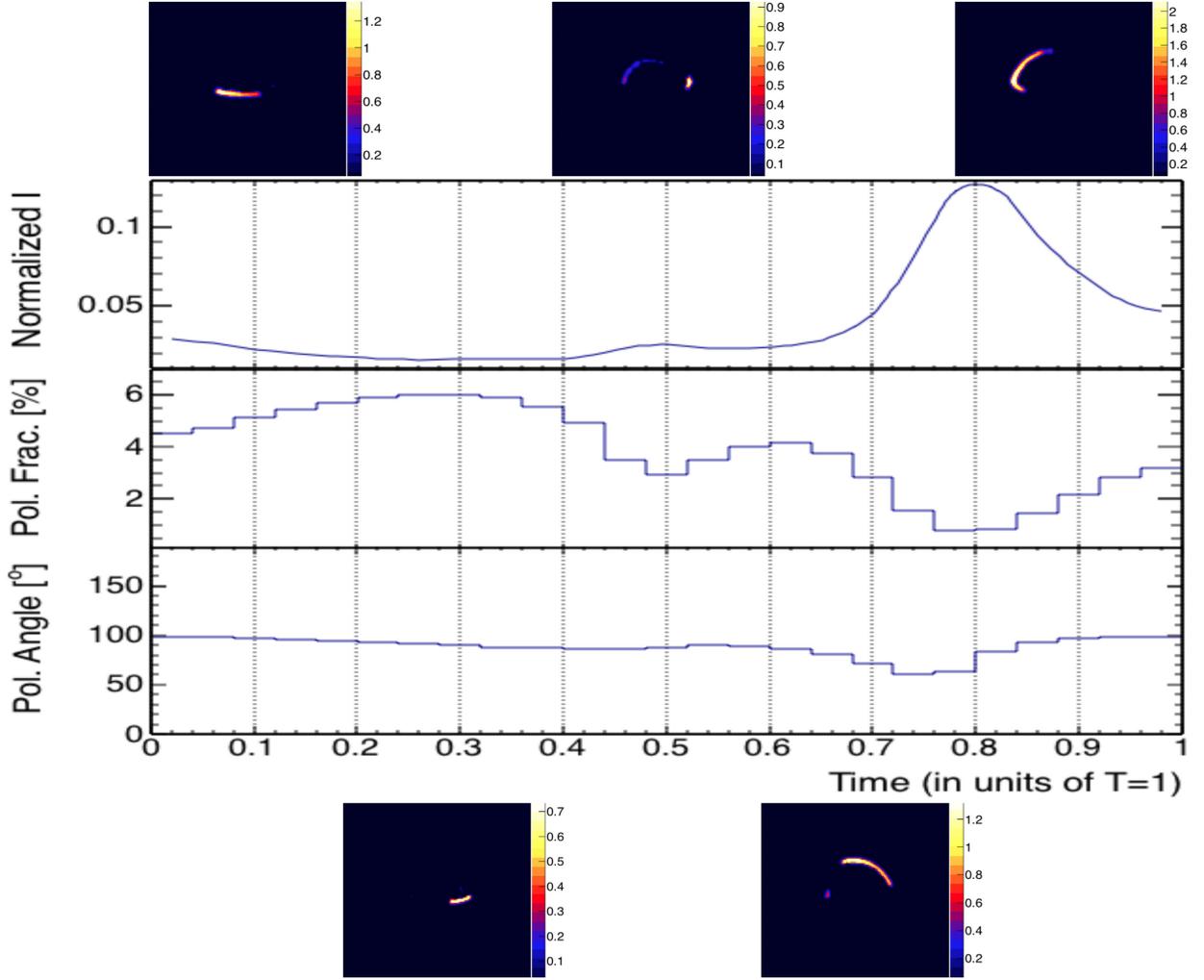} 
\caption{Light curve, polarization fraction and angle, and images of the direct emission from a HS orbiting the Schwarzschild BH viewed at $75^{\circ}$ inclination (relative to the spin axis of the accretion disk). The images show the HS in five phase bins. For instance, the first image (top left) shows the emission of the phase bin from $t=0$ to $t=0.2~T$, $T$ being the HS period. The axis label and scale for the images are the same as Fig.~\ref{map}. The intensity is normalized to 1 when integrated over all phases.
\newline(A color version of this figure is available in the online journal.) }\label{plotmap1}
\end{figure}

\begin{figure}
\epsscale{1.1}
\plotone{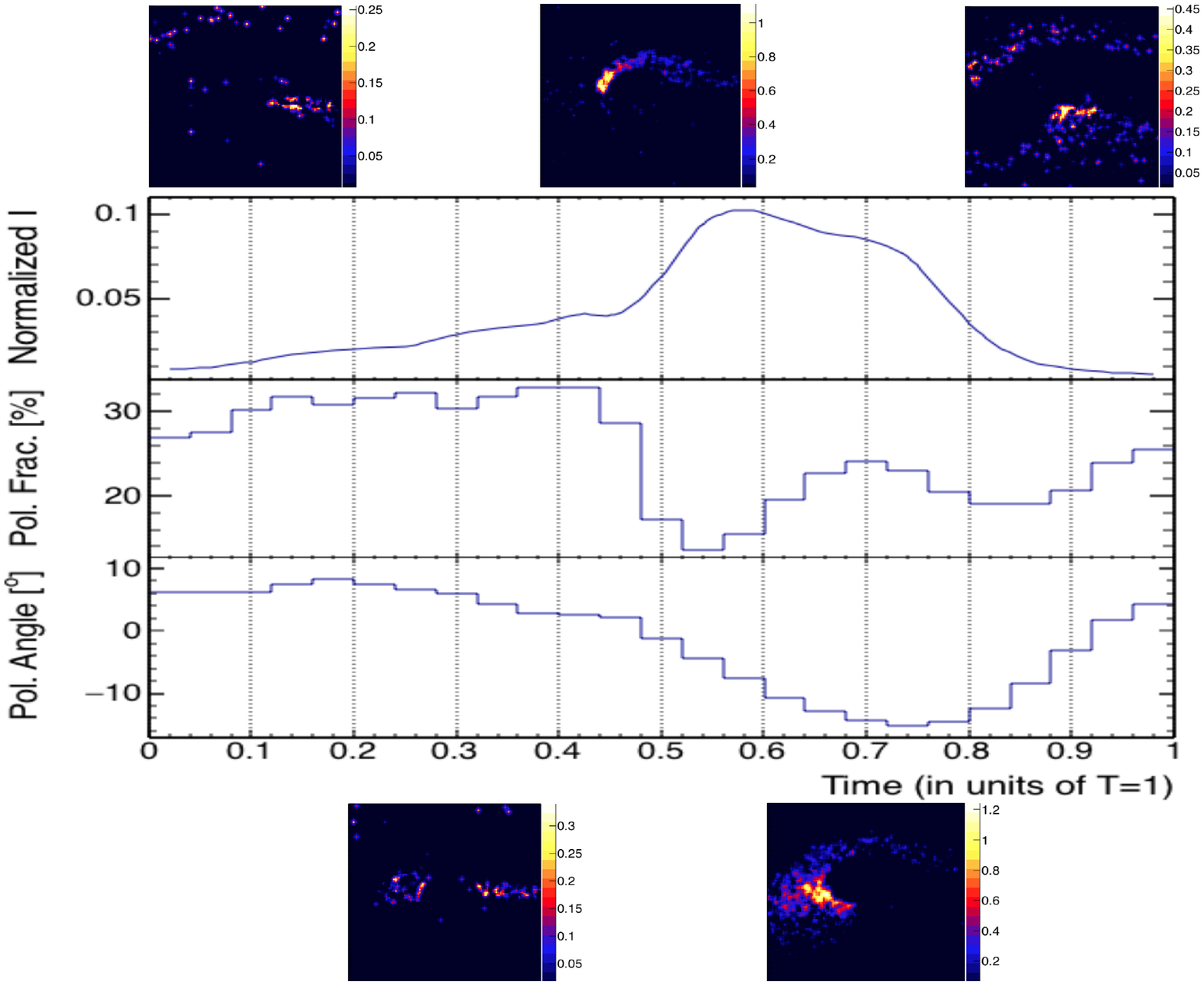} 
\caption{The same as Fig.~\ref{plotmap1} for the emission returning to the accretion disk and being scattered at least once.
\newline(A color version of this figure is available in the online journal.)}\label{plotmap2}
\end{figure}

Figures~\ref{plotmap1} and~\ref{plotmap2} show (for the Schwarzschild BH) the light curve and polarization angle of 
an orbiting HS together with snapshot of images of the emission made with direct (non-scattered photons) 
and returning (scattered photons) radiation, respectively. 
In the top middle snapshot in Fig.~\ref{plotmap1} the bottom ring is observable due to the extreme curvature of the spacetime close to the BH. 
The light curve in figure~\ref{plotmap1} demonstrates the HS is brightest in the $0.7\,T-0.9\,T$ phase bin (with $T$ being the orbital period of the HS). The apparent brightness distribution results from the combined effect of relativistic boosting and light travel time effects. 
The spot appears to orbit faster during the first half of its orbit. The same result is seen for GRS~1915+105.

\begin{figure}
\epsscale{0.8}
\plotone{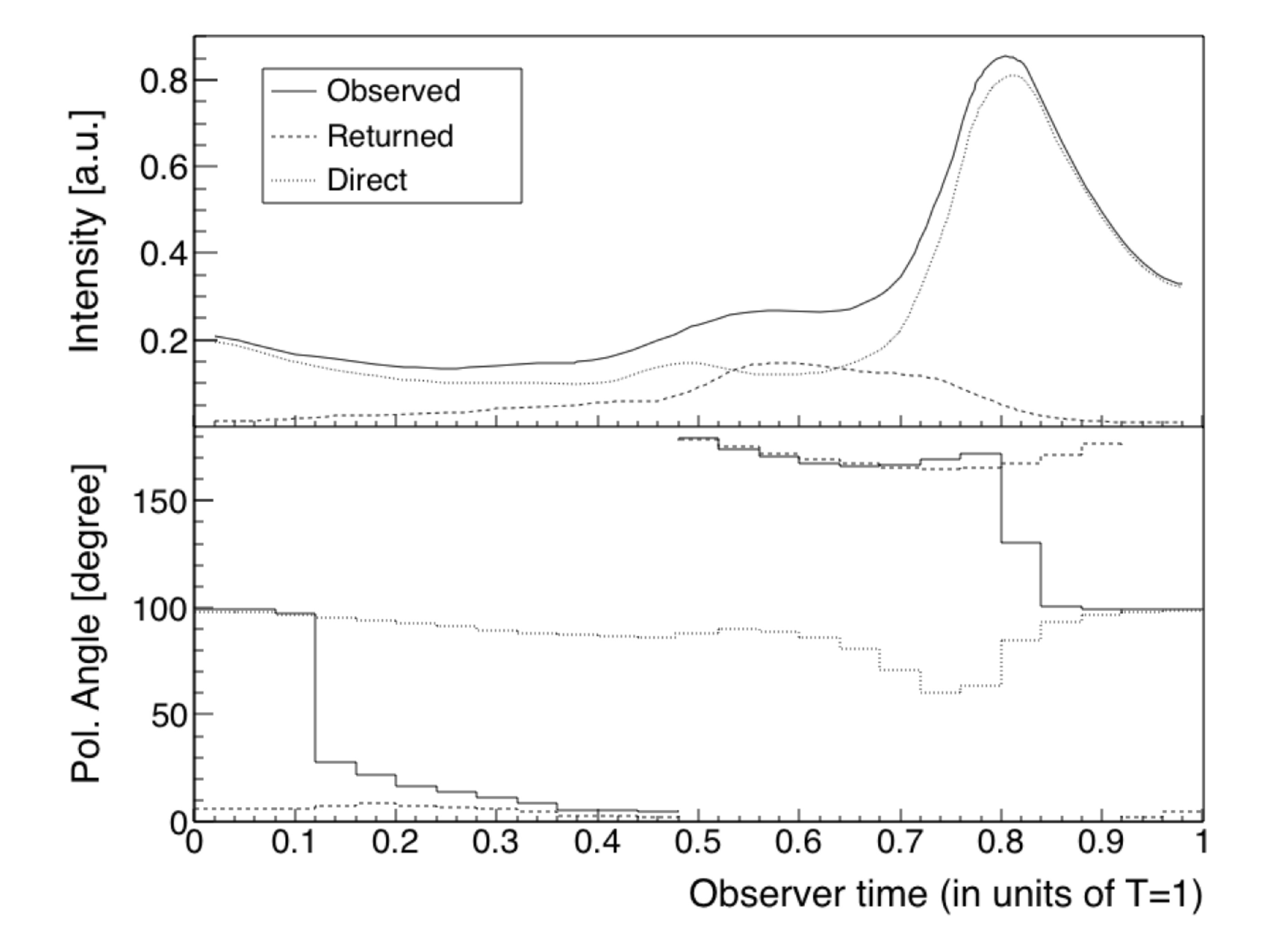} 
\caption{Intensity and polarization angle of the direct, and the returning radiation, and the sum of these (observed). 
The results show that the polarization angle is dominated by the returning radiation for the central phase bins.
}\label{plotadd}
\end{figure}

The polarization angle in figures~\ref{plotmap1} and~\ref{plotmap2} show the effect of scattering on the polarization angle. Approximately between $10-40\%$ of photons scatter off the disk depending on the phase of the HS. These scattered photons are highly polarized and thus strongly impact the net polarization of the signal (Fig~\ref{plotmap2}). The importance of the scattered photons on the polarization angle can be seen from the intensities (Fig~\ref{plotadd}). For direct photons the polarization vector is mostly parallel ($\pm 90^{\circ}$). The scattered photons acquire a 90$^{\circ}$ rotated polarization angle. In the $0.1\,T-0.7\,T$ phase bin, the returning radiation intensity becomes higher, so the observed polarization angle is dominated by returning radiation which is strongly polarized, thus it is around $0^{\circ}/180^{\circ}$. For phases over $0.8\,T$ the direct intensity with a 90 degree rotated polarization angle dominates. Furthermore, the change of the polarization angle is larger for the Kerr BH than for the Schwarzschild BH owing to the stronger curved spacetime of the Kerr BH and the larger fraction of photons returning to the accretion disk for a Kerr BH with a smaller ISCO.

All polarization plots show an anti-correlation between the intensity and the polarization fraction of the HS. 
For example, in Fig.~\ref{Ipol9}, we see that the high fluxes in the second half of the 
orbit are polarized to a low degree. The effect is smaller for the Schwarzschild BH which shows 
higher polarization fractions than the Kerr black hole at the end of the orbit. The effect of photons 
returning to the accretion disk on the polarization fraction owing to the curved space time is shown in Fig.~\ref{plotmap2}.
Not only does the figure emphasize that scattering leads to a strong polarization of the returning radiation, but also it confirms the anti-correlation of intensity and polarization fraction. The same result is seen for GRS~1915+105.

It is instructive to compare our results with those of \citet{bro05} who modeled the polarized emission of a hotspot orbiting a black hole. While the emission of the hotspots in Figs. 5 and 6 depolarizes when the intensity peaks, the hot spot emission of Broderick \& Loeb depolarizes briefly before the intensity peaks. We explain the different results by three main reasons: (i) Our code assumes that the initial polarization of the emission is given by Chandrasekhar's classical results for the emission of an optically thick atmosphere \citep{cha60}: the polarization fraction increases from zero close to the zenith to a few percent close to the horizon (where ``zenit'' and ``horizon'' refer to an observer in the disk frame) and the polarization direction is perpendicular to the plane of the zenith and the emission direction. In contrast, Broderick \& Loeb assume a constant polarization fraction, always orthogonal to the spin axis of the black hole. (ii) Whereas we modeled the X-ray emission returning to and scattering off the accretion disk (strongly impacting the observed net-polarization), Broderick \& Loeb do not do so. (iii) Broderick and Loeb assumed a different hotspot geometry and size and the predicted results are to some extend dependent on them.

\begin{figure}
\epsscale{1}
\plotone{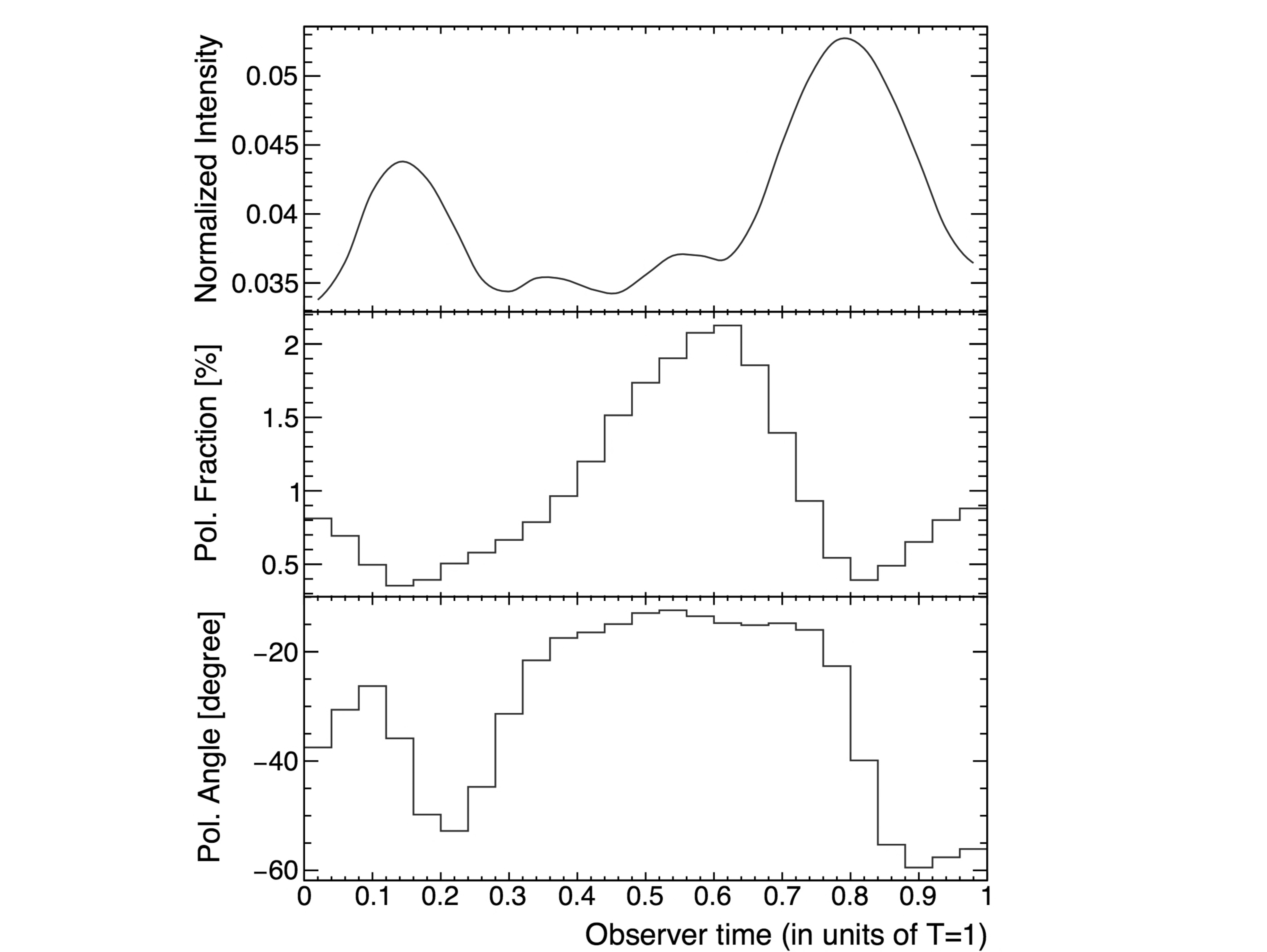} 
\caption{Intensity, polarization fraction, and polarization angle of 10 identical HSs emission for the Kerr BH, viewed at an inclination of $66^{\circ}$. The emission is polarized with a max polarization fraction of $\approx$ 2.2\%.
}\label{10S}
\end{figure}

A single pronounced hotspot produces cleaner observational signatures than a combination of several hotspots. We studied the observational appearance of multiple hotspots by simulating an accretion disk with 10 identical HSs. We assume that the HSs orbit the BH at the same distance but with a random phase. Figure~\ref{10S} shows the light curve and polarization signature of this simulation for the Kerr BH. Similar to the results for a single HS, we see that the polarization fraction anti-correlates with the flux. The polarization variation is smaller than for a single HS in the same way as a bigger HS leads to smaller polarization variations as the polarization of different parts of the HS do not add up coherently.

\begin{figure}
\epsscale{0.80}
\plotone{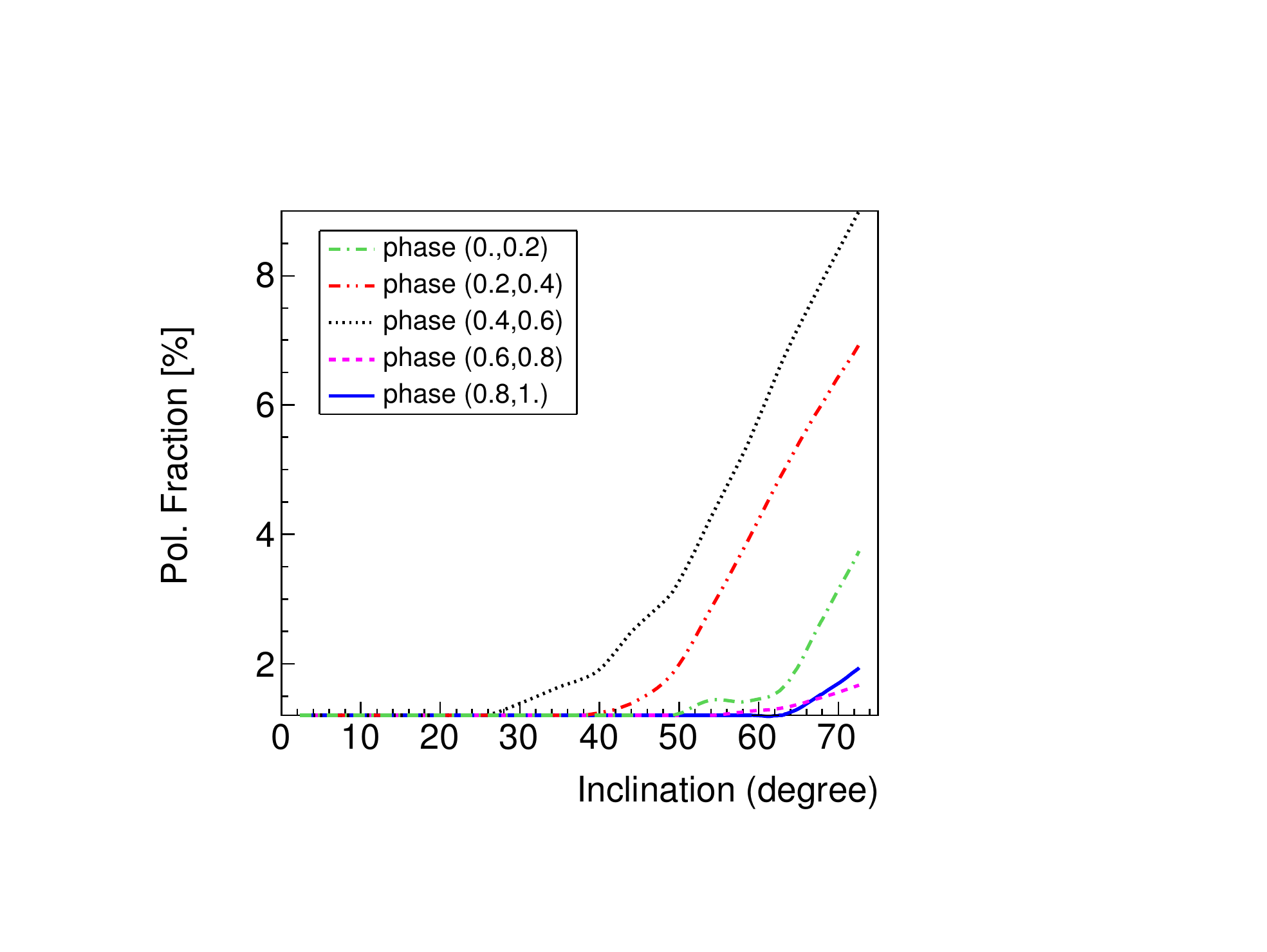}
\caption{Polarization fraction versus inclination for GRS~1915+105. Different lines show different phase bins. The polarization fractions increase with increasing inclination.
\newline(A color version of this figure is available in the online journal.) 
}\label{polincl}
\end{figure}

Furthermore, we investigate the change in polarization by changing inclination of the BH and size of the HS. Figure~\ref{polincl} shows that the polarization fraction increases with BH inclination. 
Note that in the simulation of \citet{sch04} the HFQPO amplitude exhibits a similar behavior with increasing inclination. For polarization angle, there is no simple behavior but generally it decreases by increasing inclination as a result of the lower polarization
of photons leaving the emitting plasma in its reference frame closer to the surface normal. Larger HSs are less polarized than 
smaller HSs as averaging over different polarization directions reduces the polarization fraction. 
The HS polarization also gets smaller when increasing the distance of the HS from the BH as the fraction of returning radiation decreases. We see the same result for polarization angle by enlarging the HS. Our results show that the effect of inclination and HS size on the polarization are stronger for the Kerr BH than the Schwarzschild BH. Also, in this paper we assumed that the HS temperature is $5T_{eff}$ to produce the realistic modulation in flux. Whereas the polarization of the HS is independent of its temperature, the peak energy of the emission is not. A larger HS can have a lower temperature and still produce the same flux modulation. Such a larger HS would emit less polarized emission due to averaging different polarization directions over a larger area.
\subsection{Coronal emission}
\begin{figure}
\epsscale{0.80}
\plotone{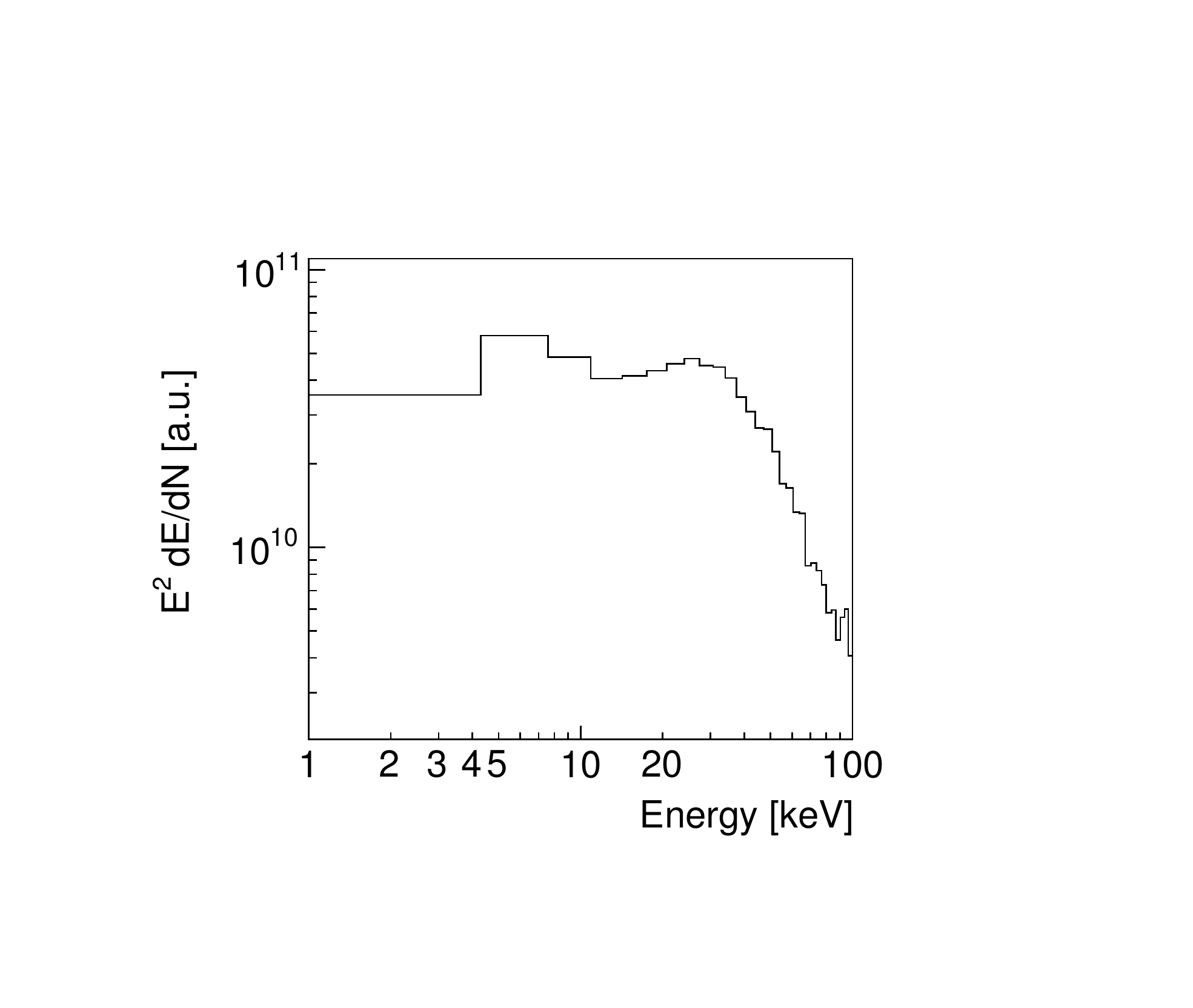}
\caption{Observed Energy flux per logarithmic energy interval $E^2 dN/dE$ from the accretion disk with a sandwich geometry for GRS 1915+105. The Comptonized spectrum has a photon index of $\approx 2.7$.}\label{fluxenergy}
\end{figure}

\begin{figure}
\epsscale{0.80}
\plotone{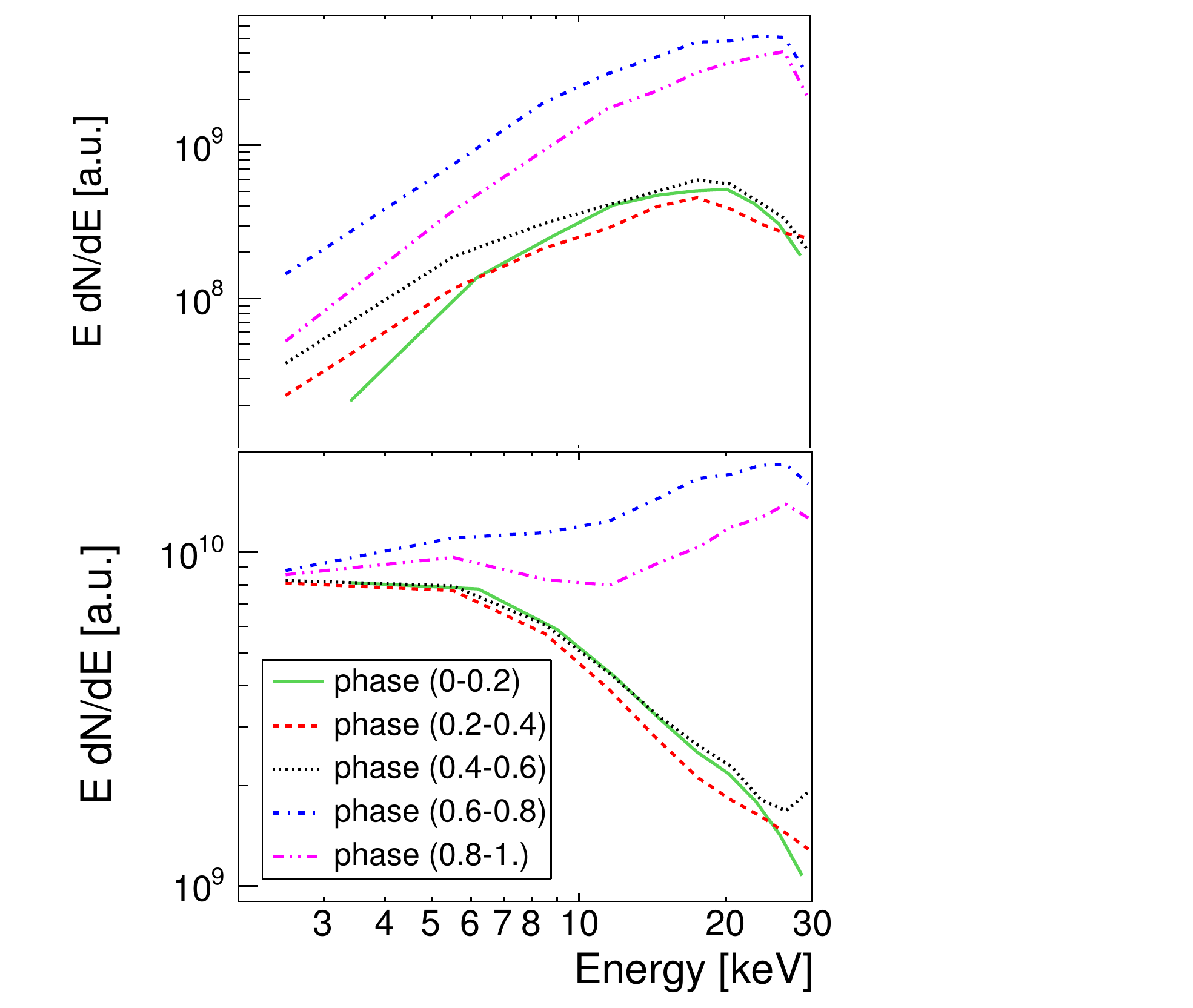}
\caption{Phase resolved energy spectra of a HS emission (upper panel) and total emission (lower panel) for GRS~1915+105 with a sandwich corona geometry.
\newline(A color version of this figure is available in the online journal.)
}\label{fluxcorona}
\end{figure}

\begin{figure}
\epsscale{1}
\plotone{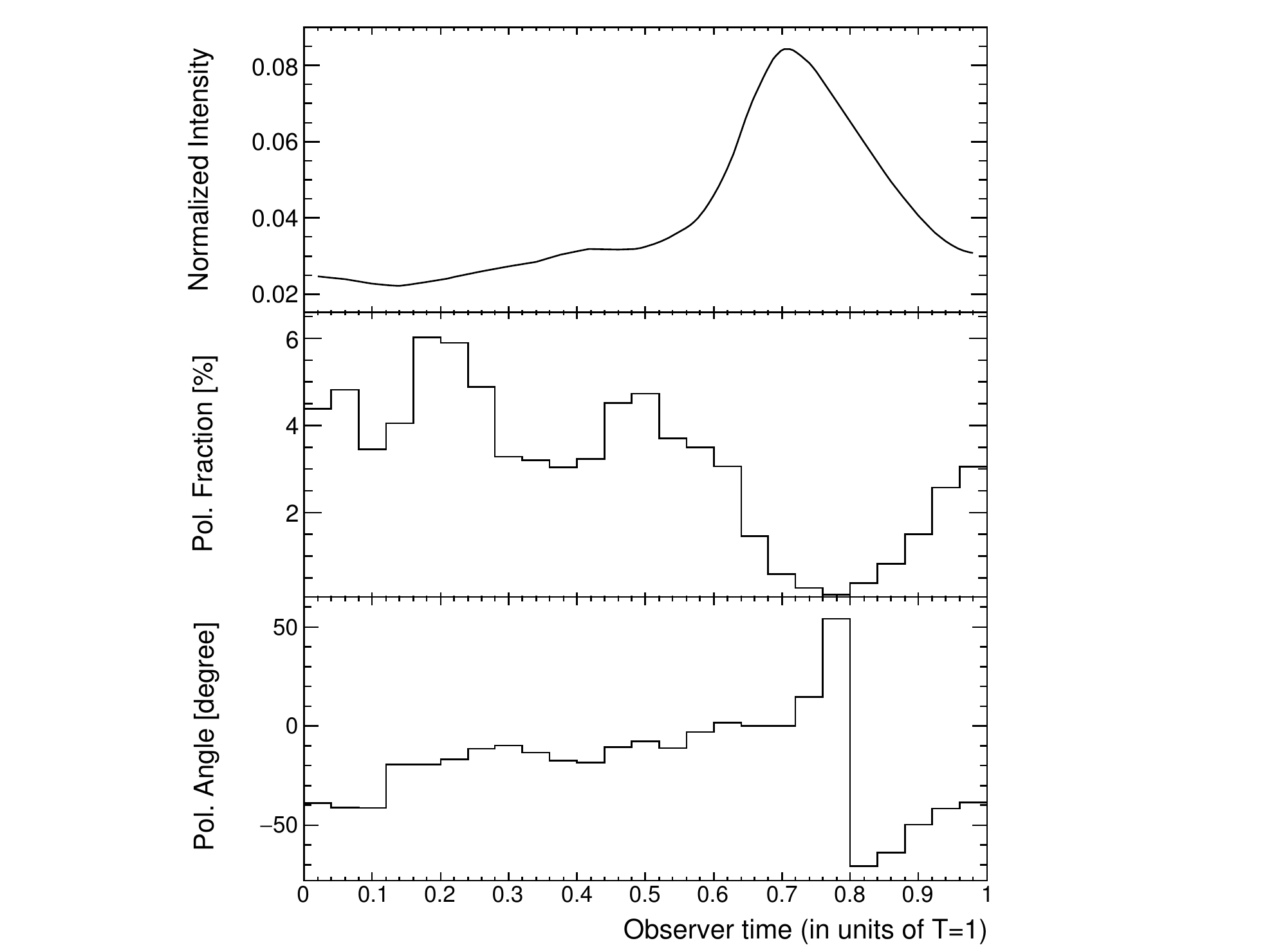}
\caption{Intensity, polarization fraction, and polarization angle of the HS coronal emission for GRS~1915+105. The emission is polarized with a max polarization fraction of $\approx$ 6\%.}\label{coronapol}
\end{figure}

Figure~\ref{fluxenergy} shows the power law tail of the observed flux for the HS and coronal emission. The simulation gives a photon index close to the one observed For GRS 1915+105 by  \citet{bel06} in the SPL state. The phase resolved energy spectra of the HS and the accretion disk are shown in Figure~\ref{fluxcorona}. The HS emission can clearly be recognized by the hard emission at the highest energies. Overall, the results look similar to the ones discussed in the absence of a corona (Fig.~\ref{flux}). Figure~\ref{coronapol} shows the normalized intensity, polarization fraction and polarization angle for the same model. Although the polarization signatures are somewhat less pronounced when accounting for the Comptonization of the emission in the corona (because of the associated light travel delays and loss of phase information), the intensity and polarization fraction still show an anti-correlation as discussed for the model without a corona. In Fig.~\ref{coronapol} the polarization peaks around the phase $0.2\,T$ where the photons scattered in the corona are more dominant.  
\section{Summary and Discussion}
This paper shows results from simulating HSs orbiting accreting Schwarzschild and Kerr BHs in X-ray binaries. 
The HS flux shows a pronounced peak accompanied by a hardening energy spectrum with the hardness peak trailing
the flux peak by 0.2 in phase. 
This specific signature could be observed by an instrument like {\it LOFT}. The mission would detect GRS~1915+105 
with a detection rate exceeding 100,000 cts/s \citep{suc12}. Using Fourier filter techniques of \citet{tom01} with the 
light curves with $>30$ detected photons during each period of the 166 Hz QPO with an rms of 6\% would make it possible to determine for each detected photon a phase.
 The phase resolved light curve would distinguish the HS model (predicting a sharp peak of
the light curve) against competing models which predict more sinusoidal variations of the flux (see the discussion below).
Phase binning the data would make it possible to determine the peak energy of the energy spectra as function of QPO phase as shown in Fig.~\ref{fluxinten}. 

\begin{figure}
\epsscale{0.8}
\plotone{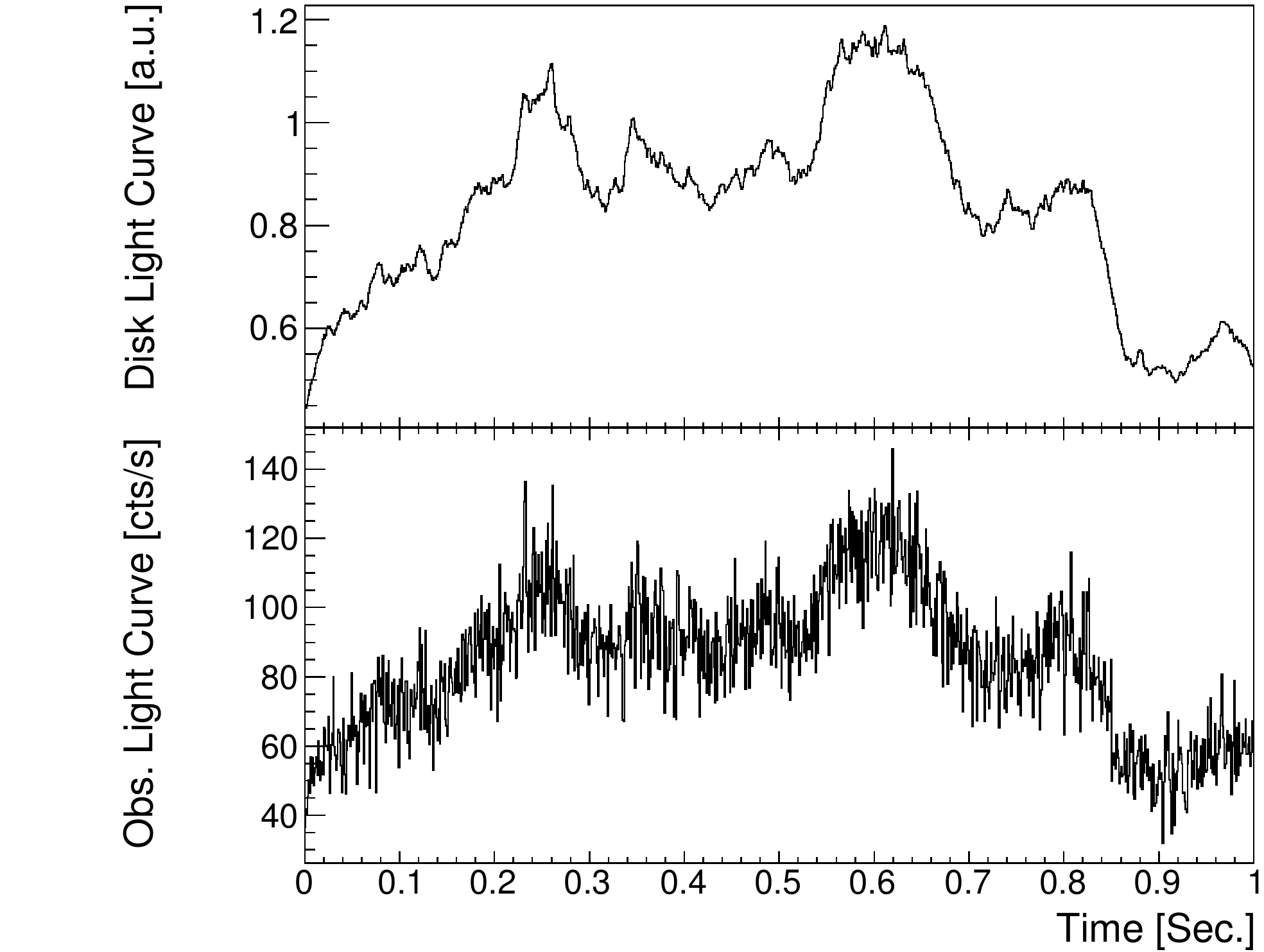}
\caption{Exemplary simulated disk emission (upper panel) and disk plus HS emission (lower panel).}\label{lcplots}
\end{figure}

\begin{figure}
\epsscale{0.90}
\plotone{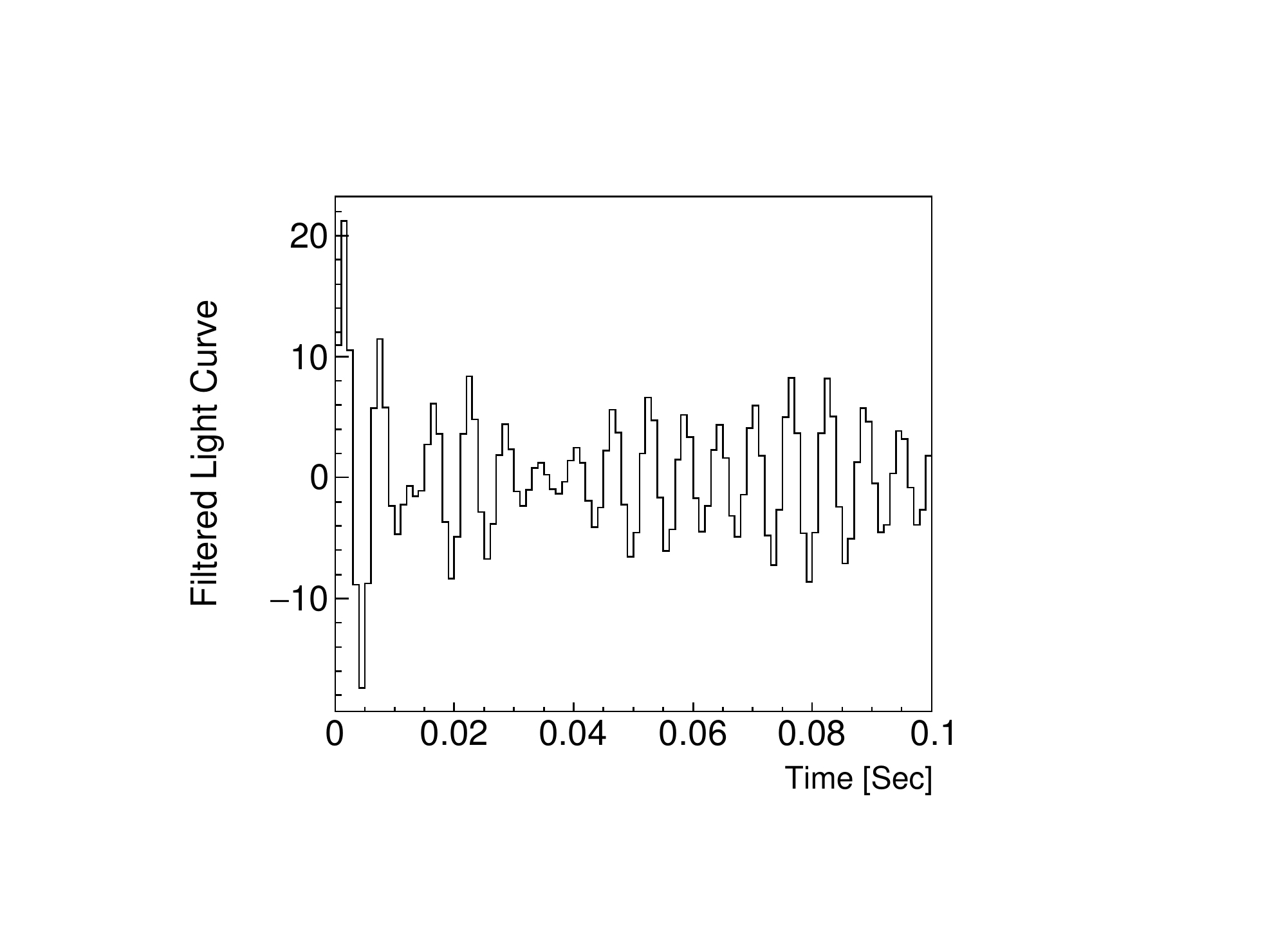}
\caption{The observed light curve predicted for  {\it LOFT} (Fig.~\ref{lcplots}) after band pass filtering.}\label{filter}
\end{figure}

\begin{figure}
\epsscale{0.90}
\plotone{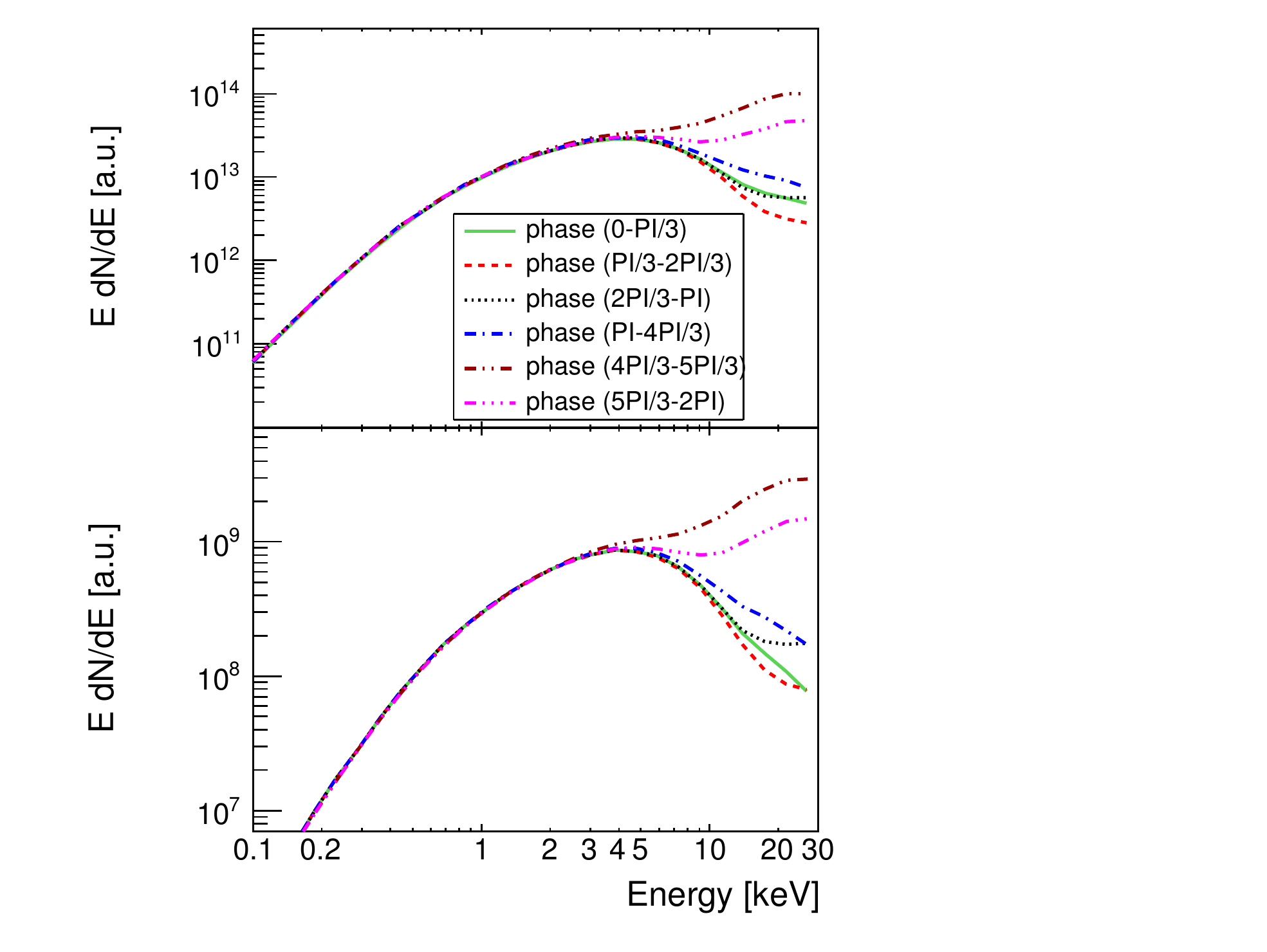}
\caption{The folded phase resolved energy spectra for the  {\it LOFT} observation simulation (upper panel) and the phase resolve energy spectra of the thermal model for total emission (lower panel). The orbital period divided in to 6 equally spaced bins each line representing the energy spectrum observed in one of these bins.
\newline(A color version of this figure is available in the online journal.)
}\label{foldflux}
\end{figure}

We carried through a detailed simulation and analysis to evaluate the detectability of the phase resolved spectral variations with  {\it LOFT}. We used the methods of  \citet{tim95} to simulate the time variable emission from the accretion disk with a realistic power spectral density (Fig.~\ref{lcplots}, top panel). We then used the methods of \citet{ing13} to simulate quasi-periodic oscillations based on the phase resolved HS intensity from Fig.~\ref{Ipol9}. Subsequently, we added statistical fluctuations to the total signal, taking the {\it LOFT} sensitivity into account. The bottom panel of Fig.~\ref{lcplots} shows the resulting light curve for a 1 sec {\it LOFT} observation. Although the long-term flux evolution is dominated by the low-frequency flux variability of the accretion disk emission, the HFQPOs with a period of $\approx 0.006s$ can clearly be recognized. Subsequently, we applied the frequency filtering method of \citet{tom01} selecting on frequencies within $\pm$20\% of the HFQPO. The filtered light curve is shown in \ref{filter}. The filtered flux curve is subsequently used to determine the {\it reconstructed phase}  . We find that the difference between the reconstructed and true phases is approximately normally distributed with a sigma of $\approx 0.08$ for a 5 minutes observation of {\it LOFT}. The phase tagging becomes more accurate as we increase the observation time. Using the reconstructed phases, we can reconstruct phase-resolved energy spectra. The lower (upper) panel of Fig.~\ref{foldflux} shows the phase resolved energy spectrum measured based on the true (reconstructed) phase information. The phase reconstruction does reduce the differences between the phase-binned energy spectra, but not catastrophically. Although we show the results here only for the HS of the thermal accretion disk, it is clear that a similar analysis could be carried through for the corona HS.  A mission like {\it LOFT} would thus make it possible to test the predictions of the HS in a good detail.
The high statistical accuracy of the data would even enable constraining the parameters 
of the hotspot (e.g. the size of the hotspot). 

The HS thermal emission (direct and reflected) is polarized to between $\sim$1\% and $\sim$10\% and exhibits large amplitude
polarization swings (see Table~\ref{table1}). According to our simulation, the HS contributes a fraction of $f \approx 9\% $ to the total emission, the HS model thus predicts that the overall polarization fraction varies by $\sim \pm f*(\Pi_{\rm max}-\Pi_{\rm min})/2 \approx 0.4 \%$ function of HS phase where $\Pi$ is the polarization fraction. This prediction for HS in coronal emission with the higher $f$ but the lower polarization variation is 0.3\%. A specific prediction of the HS model is an anti-correlation of the polarization fraction as function of the HS flux. 
The polarization fraction variations of the competing HFQPO models are most likely much smaller.
In the resonance model, e.g. \citet{abr01} and \citet{abr03b}, a perturbation excites oscillatory modes close to ISCO. 
\citet{pet08} models the HFQPOs of GRS~1915+105 assuming 
a spiral wave in the inner part of the accretion disk being in resonance with vertical epicyclic oscillations. 
In this model, the brightening disk portion is a ring segment rather than a more localized HS. 
The polarization of the emission from the bright ring segment will be more similar to that of the HS {\it averaging over all phases}.
The averaging process reduced the expected polarization by a factor of a few.
In the torus model \citep{rez03} HFQPOs are the result of p-mode (pressure mode) oscillations of an 
accretion torus orbiting the BH close to the ISCO. The model assumes a non-Keplerian geometrically 
thick disk resembling a torus rather than a disk. The HFQPOs are thought to arise from hydrodynamic or
magnetohydrodynamic instabilities \citep{rez03}.  The authors set an upper limit on the radius $r_{\rm t}$ 
of the torus of GRS~1915+105 of $r_{\rm t}<~2.7 r_g$ as in the absence of stabilizing magnetic fields, 
a larger torus would be susceptible to non-axisymmetric perturbations. We estimated the polarization 
of the emission from such a torus by considering the emission from a ring at a radial coordinate of $r_{\rm t}=2.7~r_g$. 
The ring is optically thick and for simplicity we assume that its flux changes sinusoidally with a frequency equal to the HFQPO with
a maximum flux exceeding the minimum flux by a factor of 5. The torus model predicts $\ll$1\% polarization variations. 
Furthermore, the minute peaks of the polarization fraction are {\it in phase} with the brightness peaks. 
The results described in this paragraph are summarized in Table~\ref{table2}. 
 
Could a next-generation space-borne X-ray polarimeter like {\it PolSTAR} (a space-borne version of the balloon borne X-Calibur experiment \citep{bei12, bei14, guo13} with excellent sensitivity in the $3-50$ keV energy band), {\it PRAXYS} \citep{jah15}, {\it IXPE} \citep{wei14} or {\it XIPE} \citep{sof13} detect the polarization variation predicted by the HS model? 
We considered two methods to search for the polarization variations: (i) the analysis of the Fourier transformed Stokes parameters and derived quantities,  and (ii) the analysis of the polarization fraction and angle as function of QPO phase. We evaluated the first method based on the Stokes parameters $I_i$, $Q_i$ and $U_i$ for each detected X-ray photon, as defined in \citep{kis15}.  We calculated the polarization fraction $\pi_{\rm k}$ of the $k^{\rm th}$  time bin with the standard equation: $\pi_{\rm k}=\sqrt{\sum_{i_k} Q_{i_k}^2+U_{i_k}^2}/\sum_{i_k} I_{i_k}$. However, the Fourier transform of  $\pi_{\rm k}$ did not show pronounced peaks near the QPO frequency, indicating that quantities other than $\pi_{\rm k}$ should be used to search for quasi-periodic variations of the polarization fraction. The second method requires us to determine a phase for each individual detected event, 
enabling the determination of phase-binned polarization fractions and polarization angles. 
As the GRS~1915+105 detection rate of first-generation polarimeters would be $\sim$100 cts/s, they would detect less than 
one photon during each HFQPO cycle (and an even smaller fraction of HS photons). Such a low rate would 
not enable the assignment of a QPO phase. The study of the polarization properties of HFQPOs would thus 
require the concurrent operation of a first generation X-ray polarimeter with a {\it LOFT}-type timing mission.
The latter instrument would supply the information for phase binning the data from the polarimeter mission.
Whereas the systematic errors on absolute polarization fraction measurements with a polarimeter like {\it PolSTAR} 
are on the order of $0.25\%$, the systematic errors on {\it short term polarization fraction variations} are much smaller. 
We conclude that the detection of the hotspot polarization signatures would be challenging 
but not entirely impossible.

In this paper we simulated the simple thermal disk and a wedge corona geometry with a HS to model spectral and polarization signatures of a HS.  Other disk models like ADAF can produce a very hot gas in the innermost region of the disk, making the HS with temperature higher than 5 keV which can produce seed photons that are already in high energy bands, with a moderate up scattering in the small coronal region \citep{sch06}. Also it will be exciting to do similar studies based on GRRMHD codes which evolve the accretion disk and a hotspot self-consistently.
\section{Acknowledgments}
The authors acknowledge NASA support under grant \#~NNX14AD19G.
BB acknowledges fellowship support through the Graduate School of Arts \& 
Sciences of Washington University in St. Louis. 
The authors thank A. Ingram and F. Kislat for valuable comments.
\acknowledgments

\clearpage

\clearpage

\begin{deluxetable}{ccrrrrrrrrcrl}
\tabletypesize{\scriptsize}

\tablecaption{Polarization properties of the Schwarzschild and Kerr BHs\label{table1}}
\tablewidth{0.pt}
\tablehead{
\colhead{Black hole} & \colhead{ Hotspot Min polarization fraction} & \colhead{ Hotspot Max polarization fraction} & \colhead{Disk polarization fraction}}
\startdata
Schwarzschild (spin$=0$) & 0.17 \% & 8.4 \% & 3.2 \%\\
Kerr (spin$=0.95$) & 0.21 \% & 9.5 \% & 1 \% \\
\enddata
\end{deluxetable}

\begin{deluxetable}{ccrrrrrrrrcrl}
\tabletypesize{\scriptsize}

\tablecaption{Polarization properties of different HFQPO models\label{table2}}
\tablewidth{0pt}
\tablehead{
\colhead{Model} & \colhead{Refrence} & \colhead{Average polarization fraction} & \colhead{Max range of pol. frac. variation}
}
\startdata
HS model & \citet{sch04} & 4.86 \% & 4.6 \% \\
Resonance model & \citet{pet08} & 0.78 \% & $<0.1$ \%\\
Torus model & \citet{rez03} & 2.97 \% & $<0.1$ \%\\
\enddata

\end{deluxetable}

\end{document}